\documentclass[preprint,12pt]{elsarticle}

\usepackage{amssymb}
\usepackage{amsmath}

\usepackage{placeins}
\usepackage{caption}
\usepackage{subcaption}
\usepackage{lineno}
\usepackage{xcolor}
\usepackage{booktabs}
\usepackage{multirow}
\usepackage{url}
\usepackage{hyperref}
\usepackage{algorithm}
\usepackage{booktabs}
\usepackage{siunitx}
\usepackage{algpseudocode}

\journal{Journal of Computational Physics}
\begin{document}
\begin{frontmatter}

\title{Robust Matrix-Free Newton-Krylov Solvers via Automatic Differentiation}
\author{Marco Pasquale} 
\author{Stefano Markidis} 

\affiliation{organization={Department of Computational Science and Technology, KTH Royal Institute of Technology},
            addressline={Lindstedtsvägen 5}, 
            city={Stockholm},
            postcode={11428}, 
            country={Sweden}}

\begin{abstract}
Jacobian-Free Newton-Krylov (JFNK) methods avoid forming the full Jacobian, but still require Jacobian-vector products, i.e., Gateaux derivatives of the nonlinear residual along Krylov directions. In standard Finite Differences (FD) formulations, these products are obtained by perturbing the Newton state and differencing residuals, making the linearization sensitive to round-off error and floating-point precision. This work evaluates the global impact of forward-mode Automatic Differentiation (AD) as a replacement for FD Jacobian-vector product in finite-precision JFNK solvers. The comparison keeps the discretization, Newton iteration, line search, Krylov methods, tolerances, and CPU/GPU backend fixed, only varying linearization strategy. Benchmarks include Burgers dynamics, Su-Olson radiation diffusion, reaction-diffusion, and nonlinear time-harmonic Maxwell equations, each evaluated in different nonlinear regimes. By preventing degradation of the Krylov operator, AD accelerates computation by 2–3 orders of magnitude across both CPU and GPU architectures. More importantly, it drastically improves global solver robustness, achieving a minimum completion rate of 95\%, compared to just 42\% for FD. Ultimately, accurate Gateaux derivatives unify performance and accuracy in JFNK methods, making AD the optimal choice for stiff nonlinear and reduced-precision environments.
\end{abstract}



\begin{keyword}


Jacobian-Free Newton Krylov
\sep
Automatic Differentiation
\sep
Jacobian-vector products
\sep
Finite-precision arithmetic
\sep
Nonlinear PDEs
\end{keyword}

\end{frontmatter}




\section{Introduction}\label{sec:intro}
The solution of nonlinear problems is central to numerous scientific and engineering applications, making robust nonlinear solvers indispensable in computational physics. In particular, Jacobian-Free Newton-Krylov~(JFNK) algorithms are extensively employed for large-scale systems governed by Partial Differential Equations~(PDEs). These methods are advantageous because they couple the rapid convergence characteristics of Newton's method with the memory efficiency of matrix-free Krylov subspace linear solvers~\cite{knoll_jacobian-free_2004,kelley_1995}. Additionally, JFNK provides a highly efficient computational framework capable of addressing both implicit time-dependent initial boundary value problems~(IBVPs) and boundary value problems~(BVPs). Instead of forming and storing the full Jacobian matrix, the Newton correction is computed using Krylov subspace methods such as GMRES~\cite{saad_gmres_1986}, Conjugate Gradient (CG) \cite{hestenes_methods_1952}, or BiCGSTAB~\cite{van_der_vorst_bi-cgstab_1992}. These methods require only the action of the Jacobian on vectors. This matrix-free formulation avoids explicit Jacobian assembly, but it does not avoid linearization. At each Newton iteration, the Krylov solver requires repeated Jacobian-vector products, which correspond to Gateaux derivatives of the nonlinear residual along Krylov directions. The quality of these products directly affects the Krylov subspace, the computed Newton correction, and therefore the convergence of the full nonlinear solve. For this reason, the Jacobian-vector product is a critical numerical component of the JFNK method which links accuracy to performance.

In standard JFNK implementations, Jacobian-Vector Products~(JVPs) are commonly approximated by Finite Differences~(FD) of the nonlinear residual~\cite{knoll_jacobian-free_2004,pernice_nitsol_1998}. This requires perturbing the current Newton state along a Krylov direction and differencing two residual evaluations. The perturbation size must be selected by the user or by a heuristic rule. However, its choice might have large implications. If the perturbation is too large, the derivative is biased by truncation error. If it is too small, cancellation and round-off might dominate. This sensitivity becomes especially important when using reduced precision, e.g. single-precision (FP32) instead of double-precision (FP64). In these cases, the admissible range of perturbation sizes narrows and inaccurate JVP can lead to Krylov stagnation, poor Newton corrections, and solver failure. The broader question of when derivatives should be trusted over finite differences has also been studied in the context of noisy computational functions~\cite{more_you_2014}.

The increased use of reduced-precision arithmetic, spurred by Machine Learning (ML) workloads and accelerators with FP32 and half-precision (FP16 units), makes this issue increasingly important. In several areas of computational science, including computational fluid dynamics, weather and climate modelling, and accelerator-based simulation, single- and mixed-precision are attractive because they reduce memory traffic and can improve throughput on modern hardware \cite{lang_more_2021,hatfield_improving_2018,freytag_impact_2022,karp_effects_2026,siklosi_reduced_2026}. However, the benefit of reduced precision is not automatic for nonlinear solvers. The inner Krylov iteration can be affected by round-off error, loss of orthogonality in the Arnoldi or Lanczos process, and inaccurate FD Jacobian-vector products~\cite{giraud_loss_2005}. Consequently, the linearization strategy can determine whether a JFNK solver actually benefits from reduced precision.

Automatic Differentiation~(AD)~\cite{sapienza_differentiable_2025} provides an alternative to FD for computing derivatives of numerical algorithms~\cite{corliss_automatic_2002,griewank_algorithm_1996,hascoet_tapenade_2013}. AD has been used to compute sparse Jacobians and derivative matrices in scientific computing~\cite{coleman_efficient_1998,xu_efficient_2013}, and modern tools now support increasingly complex software environments and parallel programming models~\cite{moses_scalable_2022}. AD has also been incorporated into scientific computing frameworks and multiphysics environments, including PETSc-based Newton--Krylov workflows, MOOSE, and Trilinos toolchains~\cite{hovland_parallel_2001,lindsay2021automatic,TrilinosPaper}. More recently, differentiable programming has renewed interest in differentiating numerical solvers for differential equations, especially in inverse problems, data assimilation, and physics-informed learning~\cite{raissi_physics-informed_2019,cho_separable_2023,markidis_old_2021}. Although AD offers advantages, JFNK solvers have been traditionally implemented with FD to avoid coding complexity and performance overheads associated with it. In this work, we use AD as a numerical device within a matrix-free Newton--Krylov method. Forward-mode AD computes the Jacobian-vector product required by the Krylov solver by directly differentiating the implemented discrete residual.

Replacing FD JVPs with AD alters the linearized operator presented to the Krylov solver. JFNK tightly couples multiple iterative layers, including the outer continuation loop, Newton iterations, Krylov subspace construction, and line search. Because of this nested integration, errors in the JVP can easily cascade. Therefore, the impact of AD must be evaluated across the complete nonlinear solver, not just as an isolated derivative.

The objective of this paper is to provide a complete evaluation of forward-mode AD as a replacement for FD JVP in JFNK solvers for nonlinear PDEs. We measure how this choice affects Krylov convergence, Newton convergence, execution time, and solver failure rates, at the same time comparing any possible additional overhead introduced by AD. In this work, robustness refers to the ability of the full nonlinear solver to converge reliably across problem classes, parameter regimes, floating-point precisions, and Krylov methods without stagnation or problem-specific tuning.

Our benchmark suite covers both time-dependent IBVPs and frequency-domain BVPs. It includes viscous Burgers dynamics, Su-Olson radiation diffusion, reaction-diffusion, and nonlinear time-harmonic Maxwell equations in Kerr media. These problems span nonsymmetric and symmetric systems, in parameter regimes with different degrees of stiffness. We evaluate FP64 and FP32 arithmetic, CPU and GPU execution, and multiple Krylov solvers. The goal is to identify how Jacobian-vector product construction affects the reliability and performance of practical matrix-free nonlinear PDE solvers. We demonstrate that AD improves robustness primarily by preventing finite-precision degradation of the Krylov operator. The largest differences appear in FP32, where FD JVP can cause severe Krylov stagnation. In a representative FP32 Burgers case, FD approximations inflated the iteration count to near the GMRES cap, requiring hundreds of thousands of total Krylov iterations to reach convergence. By comparison, AD maintained a tightly bounded Krylov subspace, reducing the iterations by 491$\times$ and accelerating the full nonlinear solve by 169$\times$, recovering the same converged solution.

\section{Preliminaries}\label{sec:preliminaries}

\noindent We consider a continuous nonlinear PDE problem generally written as
\begin{equation}
    \mathbf{G}\left(x_1,\dots,x_n,u,
    \frac{\partial u}{\partial x_1},\dots,
    \frac{\partial u}{\partial x_n},
    \frac{\partial^2 u}{\partial x_i\partial x_j},\dots\right)=0,
\end{equation}
where $u = u (x_1, \dots, x_n)$ is the state variable and $\mathbf{G}$ the nonlinear differential operator. After spatial discretization, and time discretization when applicable, the problem is written as a nonlinear residual equation
\begin{equation}
    \mathbf{F}(\mathbf{x}^k,\mathbf{x}^*)=\mathbf{0},
\end{equation}
where $\mathbf{x}^k$ is the current Newton iterate and $\mathbf{x}^*$ is the previously converged state or initial guess used to define the current nonlinear solve. At each Newton iteration, the residual is linearized around $\mathbf{x}^k$. Omitting the explicit dependence on $\mathbf{x}^*$, the first-order expansion gives
\begin{equation}
    \mathbf{F}(\mathbf{x}^{k+1})
    \approx
    \mathbf{F}(\mathbf{x}^k)
    + \mathcal{J}(\mathbf{x}^k)
    (\mathbf{x}^{k+1}-\mathbf{x}^k),
\end{equation}
where \(\mathcal{J}=\mathbf{F}'\) is the Jacobian of the discrete residual. Defining the Newton correction as
$\delta \mathbf{x}^k=\mathbf{x}^{k+1}-\mathbf{x}^k$, the Newton step is obtained from
\begin{equation}
    \mathcal{J}(\mathbf{x}^k)\delta \mathbf{x}^k
    =
    -\mathbf{F}(\mathbf{x}^k).
    \label{eq:newton_linear_system}
\end{equation}
JFNK methods solve Eq.~\ref{eq:newton_linear_system} with a Krylov method without explicitly assembling $\mathcal{J}$~\cite{knoll_jacobian-free_2004,kelley_1995}. The nonlinear iteration is stopped when
\begin{equation}
    \frac{
    \|\mathbf{F}(\mathbf{x}^k,\mathbf{x}^*)\|
    }{
    \|\mathbf{F}(\mathbf{x}^*,\mathbf{x}^*)\|
    }
    <
    \mathrm{tol}_{\mathrm{Newton}}.
\end{equation}
A backtracking line search can be applied after the Krylov solve by replacing
$\mathbf{x}^{k+1}=\mathbf{x}^k+\delta\mathbf{x}^k$ with
$\mathbf{x}^{k+1}=\mathbf{x}^k+\alpha\delta\mathbf{x}^k$, where $\alpha$ is reduced until sufficient nonlinear residual decrease is obtained~\cite{pernice_nitsol_1998}.

The main operation in a matrix-free Newton-Krylov method is the Jacobian-vector product. At Krylov iteration $m$, the Krylov solver provides a direction $\mathbf{v}_m$ and requires $\mathcal{J}(\mathbf{x}^k)\mathbf{v}_m$, i.e., the Gateaux derivative of the discrete residual in the direction $\mathbf{v}_m$:
\begin{equation}
    \mathcal{J}(\mathbf{x}^k)\mathbf{v}_m
    =
    \left.
    \frac{d}{d\epsilon}
    \mathbf{F}(\mathbf{x}^k+\epsilon\mathbf{v}_m)
    \right|_{\epsilon=0}.
    \label{eq:jvp_gateaux}
\end{equation}
This product determines the linear operator seen by the Krylov solver. Consequently, errors in the Jacobian-vector product affect the Krylov subspace, the computed Newton correction, and the convergence of the full nonlinear solve. In standard JFNK implementations, Eq.~\ref{eq:jvp_gateaux} is approximated by finite differences,
\begin{equation}
    \mathcal{J}(\mathbf{x}^k)\mathbf{v}_m
    \approx
    \frac{
    \mathbf{F}(\mathbf{x}^k+\epsilon \mathbf{v}_m)
    -
    \mathbf{F}(\mathbf{x}^k)
    }{\epsilon}.
    \label{eq:jvp_fd}
\end{equation}
The perturbation size \(\epsilon\) is usually chosen using a machine-precision-based rule, for example
\begin{equation}
    \epsilon
    =
    \sqrt{\epsilon_{\mathrm{mach}}}
    \max(1,\|\mathbf{x}^k\|),
    \label{eq:fd_step}
\end{equation}
possibly with additional clipping. This choice balances truncation error against cancellation and round-off error~\cite{knoll_jacobian-free_2004}. In FP32, this balance becomes more fragile because the range of useful perturbation sizes is narrower.

Forward-mode AD provides an alternative to the FD approach. Instead of differencing two residual evaluations, AD differentiates the implemented discrete residual directly and returns the primal residual together with its directional derivative,
\begin{equation}
    \mathrm{jvp}(\mathbf{F},\mathbf{x}^k,\mathbf{v}_m)
    =
    \left(
    \mathbf{F}(\mathbf{x}^k),
    \mathcal{J}(\mathbf{x}^k)\mathbf{v}_m
    \right).
    \label{eq:jvp_ad}
\end{equation}
AD differentiates the operations present in the residual evaluation. For smooth and deterministic residuals, this provides a consistent matrix-free linearization up to floating-point arithmetic. If the residual contains nonsmooth switches, adaptive discontinuities, or noisy embedded solvers, AD may differentiate these algorithmic artifacts. In such cases, carefully chosen finite differences can sometimes act as a macroscopic smoothing mechanism~\cite{more_you_2014}. The present work focuses on the regime where the dominant difficulty is the finite-precision degradation of FD Jacobian-vector products.

For a fixed Newton iterate, Krylov methods approximate the solution of Eq.~\ref{eq:newton_linear_system} in the subspace
\begin{equation}
    \mathcal{K}_m
    =
    \operatorname{span}
    \left\{
    \mathbf{r}_0,
    \mathcal{J}\mathbf{r}_0,
    \mathcal{J}^2\mathbf{r}_0,
    \dots,
    \mathcal{J}^{m-1}\mathbf{r}_0
    \right\},
    \label{eq:span}
\end{equation}
where
$\mathbf{r}_0=-\mathbf{F}-\mathcal{J}\mathbf{v}_0$ is the initial linear residual. The Krylov solver therefore repeatedly queries the Jacobian-vector product. Any noise or bias in this oracle changes the effective Krylov operator and can lead to stagnation or inaccurate Newton corrections.

The Krylov method must be chosen according to the properties of the linearized problem. CG is efficient and uses short recurrences, but it is appropriate only for symmetric positive-definite systems~\cite{hestenes_methods_1952}. GMRES is more generally applicable to nonsymmetric systems and minimizes the residual norm over $\mathcal{K}_m$, but it requires storing an expanding orthogonal basis and is often restarted in practice~\cite{saad_gmres_1986}. BiCGSTAB also applies to nonsymmetric systems and has a lower memory footprint, but its convergence can be less monotone and more sensitive to ill conditioning~\cite{van_der_vorst_bi-cgstab_1992}.

Floating-point precision affects both the JVP and the Krylov basis construction. In FD JVPs, round-off and cancellation enter through Eq.~\ref{eq:jvp_fd}. In Krylov methods, reduced precision can also degrade~\cite{giraud_loss_2005}. Inaccurate JVP can increase Krylov iteration counts, produce poorer Newton corrections, trigger additional line-search steps, and ultimately cause nonlinear solver failure.

\section{Methodology}\label{sec:method}

\subsection{Solver}\label{subsec:solver_implementation}

\begin{figure}[t!]
    \centering
    \includegraphics[width=0.99\linewidth]{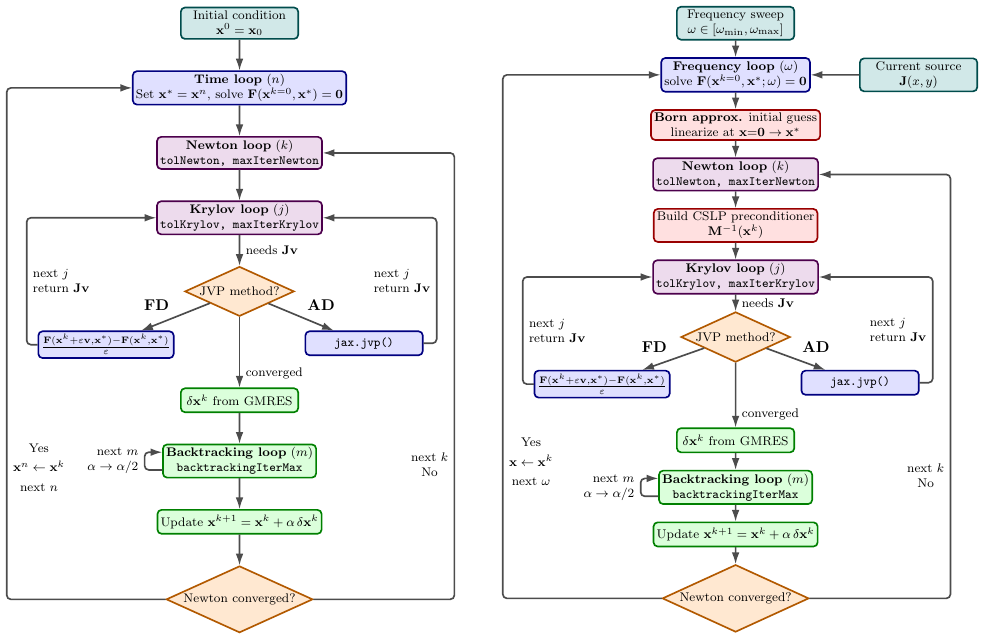}

    \hspace{5.5mm}
    \begin{minipage}{0.4\linewidth}
        \centering
        
        \footnotesize
        (a) IBVP Solver
    \end{minipage}
    \hfill
    \begin{minipage}{0.4\linewidth}
        \centering
        
        \footnotesize
        (a) BVP Solver
    \end{minipage}
    \hspace{7.5mm}

    \caption{
    Nested matrix-free Newton-Krylov workflow used in the experiments. The same outer continuation loop, time for (a) and frequency for (b), Newton iteration, Krylov solver, and line search are used for both linearization strategies. The only algorithmic difference is the JVP evaluation inside the Krylov loop, where either an FD residual difference or a forward-mode AD call is used.
    }
    
    \label{fig:solver_viz}
\end{figure}

The numerical experiments use a matrix-free JFNK solver designed to isolate the effect of the Jacobian--vector product (JVP) evaluation.  Figure~\ref{fig:solver_viz} summarizes the nested solver workflow. The outer loop advances the continuation variable: time for IBVPs and angular frequency $\omega$ for the time-harmonic Maxwell BVP. At each outer step, the nonlinear residual is solved by Newton iterations. Each Newton step calls an inner Krylov solver, which accesses the linearized problem only through matrix-free products $\mathcal{J}(\mathbf{x}^k)\mathbf{v}_m$. After the Krylov solve, a backtracking line search is applied using the same residual evaluation for both FD and AD cases.

\begin{figure}[t]
    \centering
    \includegraphics[width=0.99\linewidth, clip=true, trim=5cm 5cm 5cm 5cm]{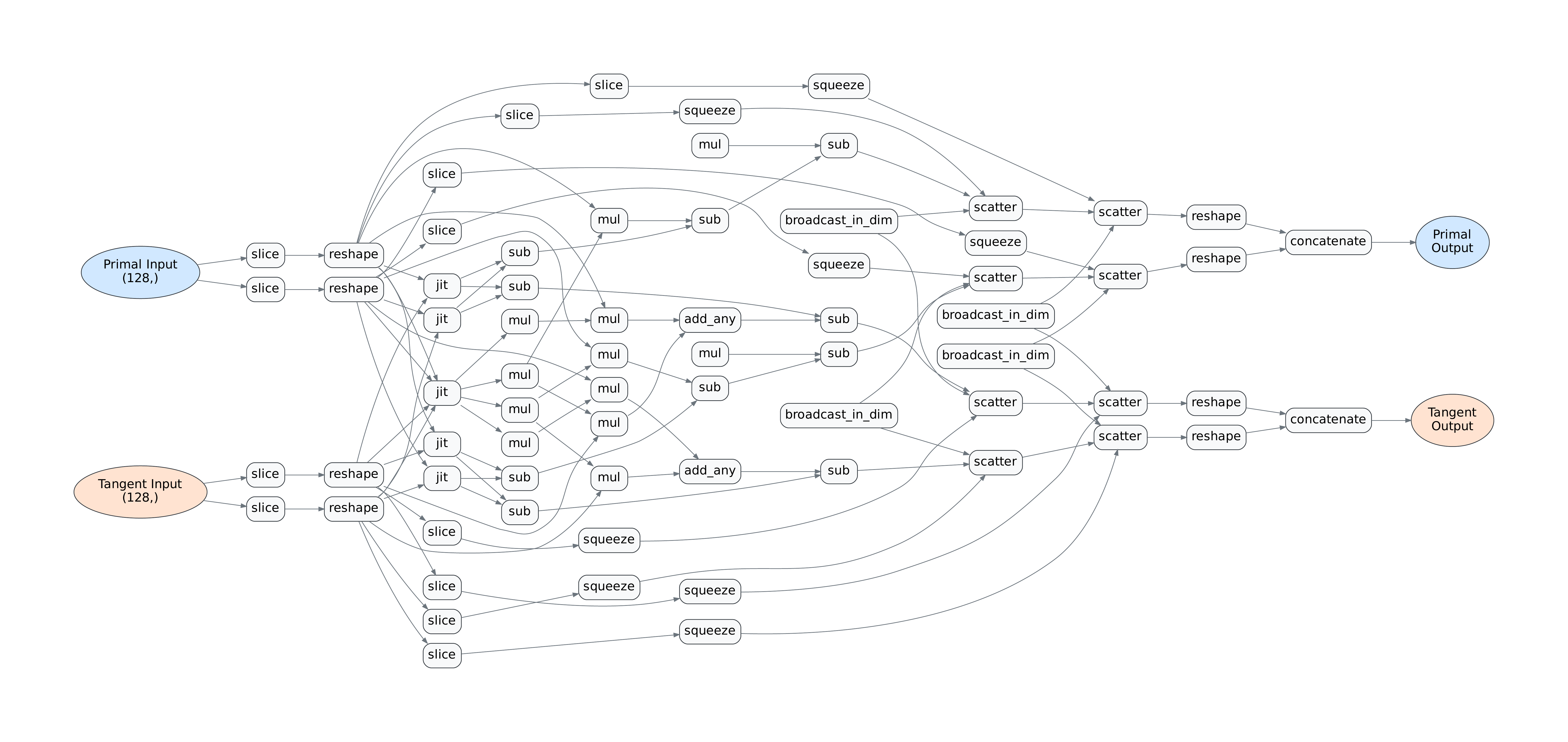}
    \caption{
    Example computational graph generated by \texttt{jax.jvp} for the time-harmonic Maxwell residual. The primal input is the current field state and the tangent input is the Krylov direction. The AD trace evaluates both the nonlinear residual and the directional derivative $\mathcal{J}(\mathbf{x}^k)\mathbf{v}_m$, which is passed to the matrix-free Krylov solver.
    }
    \label{fig:jvp_graph}
\end{figure}

The FD implementation uses the FD JVP in Eq.~\ref{eq:jvp_fd}, with the perturbation size selected according to Eq.~\ref{eq:fd_step}. The AD implementation uses the forward-mode JVP in Eq.~\ref{eq:jvp_ad}, applied to the same discrete residual function $\mathbf{F}$. In both cases, the Krylov solver receives a matrix-free linear operator. No Jacobian matrix is assembled or stored. This makes the FD and AD variants identical except for the routine that evaluates $\mathcal{J}(\mathbf{x}^k)\mathbf{v}_m$. The Krylov solvers are implemented through matrix-free linear operator interfaces. Each time the Krylov method requests a matrix-vector product, the operator calls either the FD or AD JVP routine. We use GMRES and BiCGSTAB for nonsymmetric systems, and CG when the linearized system is symmetric positive definite. This allows the effect of the JVP construction to be evaluated together with the choice of Krylov method, without changing the nonlinear solver.
The implementation combines JAX for residual evaluation and forward-mode AD with SciPy and CuPy for Krylov linear algebra. Figure~\ref{fig:jvp_graph} shows an example computational graph produced by \texttt{jax.jvp} for the Maxwell residual. The graph illustrates that the AD implementation evaluates the primal residual and the corresponding directional derivative from the same discrete residual function. This provides the matrix-free product required by the Krylov solver without assembling the Jacobian. 
On CPUs, the Krylov methods are called through \texttt{scipy.sparse.linalg} \cite{2020SciPy-NMeth}; on GPUs, the corresponding CuPy routines are used \cite{cupy_learningsys2017}. JAX provides the AD JVP through \texttt{jax.jvp} \cite{jax2018github}. We do not use JAX-native Krylov solvers because their dynamic iteration counts can lead to substantial compilation overhead. Instead, JAX evaluates the residual and AD operations, while SciPy or CuPy performs the iterative linear algebra. On GPUs, JAX and CuPy arrays are exchanged through DLPack zero-copy views, avoiding host-device transfers during the Krylov iterations.

\subsection{Benchmark Problems}\label{subsec:benchmark_problems}
The benchmark suite is designed to test how the JVP evaluation affects the robustness of the same JFNK workflow across nonlinear PDEs with different algebraic and physical structure. We consider three time-dependent initial-boundary value problems (IBVPs) and one frequency-domain nonlinear boundary value problem (BVP), spanning nonsymmetric and symmetric systems, scalar and vector unknowns, real and complex fields, and regimes with different stiffness levels. For the time-dependent problems, the residual is built from an implicit Crank-Nicolson time discretization and second-order finite differences in space; the previous time-step solution is used as the Newton initial guess. For the Maxwell problem, the continuation parameter is the angular frequency $\omega$, and the initial guess is obtained from the corresponding linearized problem. Further discretization details are given in Appendix~\ref{appdx:testcases}.

The first problem is the two-dimensional viscous Burgers equation,
\begin{equation}
    \frac{\partial \mathbf{U}}{\partial t}
    +
    (\mathbf{U}\cdot\nabla)\mathbf{U}
    =
    \nu \nabla^2 \mathbf{U},
\end{equation}
where $\mathbf{U}$ is the velocity field and $\nu$ is the kinematic viscosity. The advective nonlinearity produces nonsymmetric linearized systems, making this problem a representative test case for nonsymmetric Krylov solvers.

The second problem is the Su-Olson radiation-diffusion system \cite{castor_radiation_2004},
\begin{equation}
    \left\{
    \begin{aligned}
        \frac{\partial U}{\partial t}
        &=
        \frac{1}{3}\nabla^2 U - (U - V) + Q(\mathbf{x},t),\\
        \frac{\partial V}{\partial t}
        &=
        \xi (U - V),
    \end{aligned}
    \right.
\end{equation}
where $U$ is the radiation energy density, $V$ is the material energy density, $Q(\mathbf{x},t)$ is an imposed source, and $\xi$ controls the material response. This problem tests coupled residuals with different source configurations and material parameters.

The third problem is the scalar reaction-diffusion equation,
\begin{equation}
    \frac{\partial u}{\partial t}
    =
    D\nabla^2 u - u^3,
\end{equation}
where $D$ is the diffusion coefficient and $u^3$ is a nonlinear sink term. This case provides a structured nonlinear system in which conjugate-gradient-based solvers can also be tested when the linearized problem is symmetric positive definite.

The fourth problem is the time-harmonic Maxwell equation in a Kerr medium,
\begin{equation}
    \nabla \times \nabla \times \mathbf{E}
    -
    \omega^2\mu_0 \varepsilon(\mathbf{E})\mathbf{E}
    =
    i\omega\mu_0\mathbf{J},
\end{equation}
where $\mathbf{E}$ is the complex electric field, $\omega$ is the angular frequency, $\mu_0$ is the vacuum permeability, and $\mathbf{J}$ is the imposed current density. The field-dependent permittivity is
\begin{equation}
    \varepsilon(\mathbf{E})
    =
    \varepsilon_0(1-0.05i)\left(1+\chi|\mathbf{E}|\right),
\end{equation}
where $\chi$ controls the Kerr nonlinearity. This BVP is the stiffest benchmark in the suite because resonant regimes can strongly amplify the field and make the linearized systems ill conditioned.

\subsection{Experimental design}\label{subsec:experimental_design}
For each benchmark problem, we vary the initial condition or source configuration together with a single physical control parameter. The Burgers equation is tested using Taylor--Green vortex (TGV), double shear layer (DSL), and four-vortex collision (4VC) initial conditions with $\nu \in {0.1,,0.05,,0.01}$. The Su--Olson problem is tested using static and moving source configurations with $\xi \in {1,,0.1,,0.01}$. The reaction--diffusion problem is tested using Gaussian and sinusoidal initial data with $D \in {0.1,,0.01,,0.001}$. The Maxwell--Kerr problem is tested using Gaussian and dipole current sources with $\chi \in {0.5,,0.1,,0.05}$. Together with the grid choices described below, this parameter sweep produces 480 benchmark simulations organized into 54 nonlinear problems.

The grid sizes are selected according to the target hardware. CPU runs use \(64\times64\) and \(128\times128\) grids, while GPU runs use \(256\times256\) and \(512\times512\) grids. The time-dependent problems are advanced with Courant number \(C=1\), which gives developed nonlinear dynamics while keeping the time-integration setup fixed across all solvers. For the Maxwell--Kerr problem, a preliminary frequency sweep is used to identify strongly resonant regions. The reported runs are then restricted to frequency ranges that remain well posed while still producing stiff nonlinear solves.

Each configuration is evaluated using both JVP strategies, FD and AD, in single- and double-precision. For nonsymmetric systems, we test GMRES and BiCGSTAB. For the reaction--diffusion problem, where the linearized system is symmetric positive definite, we also include CG. This gives eight solver combinations for the nonsymmetric problems and twelve for the reaction-diffusion problem, for a total of 480 solver runs per backend.

The nonlinear and linear tolerances are chosen according to the floating-point precision to avoid both premature convergence and over-solving. In FP32, we set $\mathrm{tol}_{\mathrm{Newton}} = 10^{-4}$, $\mathrm{tol}_{\mathrm{Krylov}} = 10^{-6}$. In FP64, we set $\mathrm{tol}_{\mathrm{Newton}} = 10^{-6}$, $\mathrm{tol}_{\mathrm{Krylov}} = 10^{-8}.$ The maximum number of Newton iterations is 15 for the time-dependent IBVPs and 50 for the Maxwell BVP. The maximum Krylov iteration count is set to 100. For restarted GMRES, this corresponds to the maximum number of restart cycles. These settings are kept fixed for FD and AD so that differences in runtime, iteration count, and convergence behavior reflect the JVP evaluation rather than different solver tolerances.

\subsection{Metrics and failure criteria}\label{subsec:metrics_failure}
We  record JFNK metrics that distinguish between Krylov stagnation, poor Newton progress, and complete nonlinear failure. For each run, we collect the total execution time, total number of Krylov iterations, number of Newton iterations, residual histories, line-search activity, and convergence status. A run is considered successful if the nonlinear residual reaches the prescribed Newton tolerance for the required time steps or continuation points. A run is classified as failed if any of the following conditions is met:
\begin{enumerate}
    \item the simulation terminates prematurely because of numerical instability, overflow, non-finite values, or breakdown of the linear or nonlinear solver;
    \item for time-dependent IBVPs, more than \(20\%\) of the time steps fail to converge within two orders of magnitude of the requested Newton tolerance;
    \item for the Maxwell BVP, more than \(20\%\) of the frequency steps fail to converge to the requested Newton tolerance.
\end{enumerate}
The second criterion is included to avoid over-penalizing transient convergence difficulties in time-dependent nonlinear problems. In some strongly nonlinear regimes, especially when FD JVPs are used, a small number of time steps may exhibit poor Newton convergence while the overall simulation remains accurate and stable. Such cases are therefore distinguished from systematic solver failure.

To compare solver performance across a heterogeneous benchmark suite, we use Dolan--Mor\'e performance profiles~\cite{dolan_benchmarking_2002}. These profiles summarize both performance and robustness without allowing a small number of difficult or unusually easy cases to dominate the comparison. Let \(\mathcal{P}\) denote the set of benchmark configurations and \(\mathcal{S}\) the set of solver variants being compared. For each problem \(p\in\mathcal{P}\) and solver \(s\in\mathcal{S}\), let \(t_{p,s}\) be a performance metric, such as wall-clock time or total Krylov iterations. The performance ratio is defined as
\begin{equation}
    r_{p,s}
    =
    \frac{
    t_{p,s}
    }{
    \min_{s\in\mathcal{S}} t_{p,s}
    } .
\end{equation}
Thus, \(r_{p,s}=1\) means that solver \(s\) is the best performer for problem \(p\) according to the chosen metric. Failed runs are assigned \(r_{p,s}=\infty\), so they do not contribute to the profile at finite values of \(\tau\).The performance profile of solver \(s\) is then
\begin{equation}
    \rho_s(\tau)
    =
    \frac{1}{|\mathcal{P}|}
    \left|
    \left\{
    p\in\mathcal{P}: r_{p,s}\le \tau
    \right\}
    \right|,
    \qquad \tau\ge 1 .
\end{equation}
The value \(\rho_s(1)\) gives the fraction of benchmark configurations for which solver \(s\) is the fastest or requires the fewest iterations. As \(\tau\) increases, \(\rho_s(\tau)\) measures the fraction of problems solved within a factor \(\tau\) of the best solver. The limiting value reached at large \(\tau\) corresponds to the solver success rate over the benchmark suite.

\section{Results}\label{sec:simulationresults}

\subsection{Verification and representative solutions}
\label{subsec:verification}
\begin{figure}[t!]
    \centering
    
    \def\figH{2.95cm} 
    
    \hspace{6mm}
    \begin{minipage}{0.228\linewidth}
        \centering
        \includegraphics[height=\figH, clip = true, trim = 0.7cm 0 27cm 0cm]{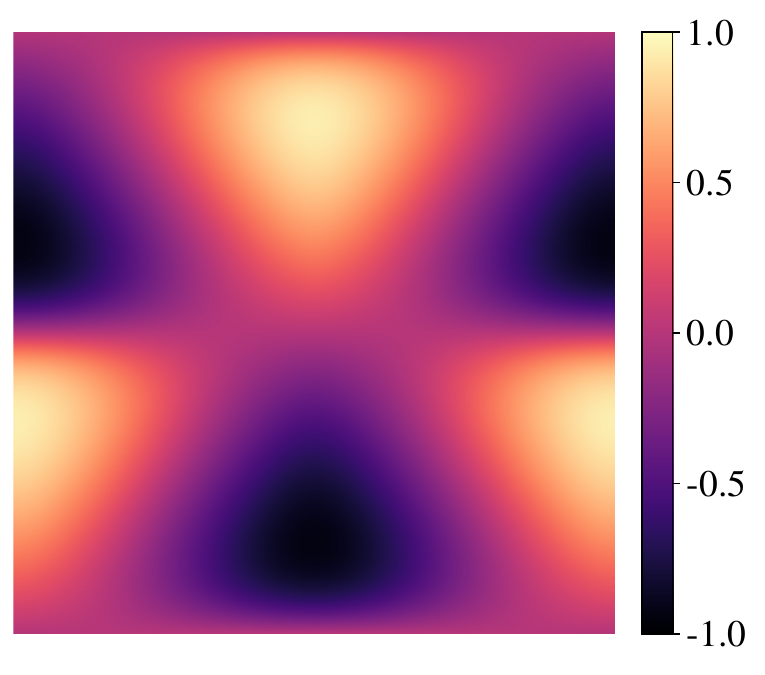}
        \scriptsize
        
        (a) $u(x,y)$
    \end{minipage}%
    \hfill
    \begin{minipage}{0.228\linewidth}
        \centering
        \includegraphics[height=\figH, clip = true, trim = 13.5cm 0 14.cm 0cm]{images/TGV1000.pdf}
        \scriptsize
        
        (b) $v(x,y)$
    \end{minipage}
    \hfill
    \begin{minipage}{0.228\linewidth}
        \centering
        \includegraphics[height=\figH, clip = true, trim = 27.5cm 0 0cm 0cm]{images/TGV1000.pdf}
        \scriptsize
        
        (c) $|\mathbf{U}(x,y)|$
    \end{minipage}
    \hspace{6mm}
    \caption{Taylor–Green vortex simulation for the Burgers' equation on a $512 \times 512$ grid after 1000 time steps. The computational domain is $x,y \in [0,2\pi)$ with fully periodic boundary conditions. The kinematic viscosity is set to $\nu = 0.05$ to induce steep velocity gradients and macroscopic shock formation. The Courant number is $\mathrm{C} = 1$.}
    \label{fig:tgv_sim}

    \vspace{4.5mm} 

    \hspace{6mm}
    \begin{minipage}{0.228\linewidth}
        \centering
        \includegraphics[height=\figH, clip = true, trim = 5.cm 1cm 25.0cm 1.3cm]{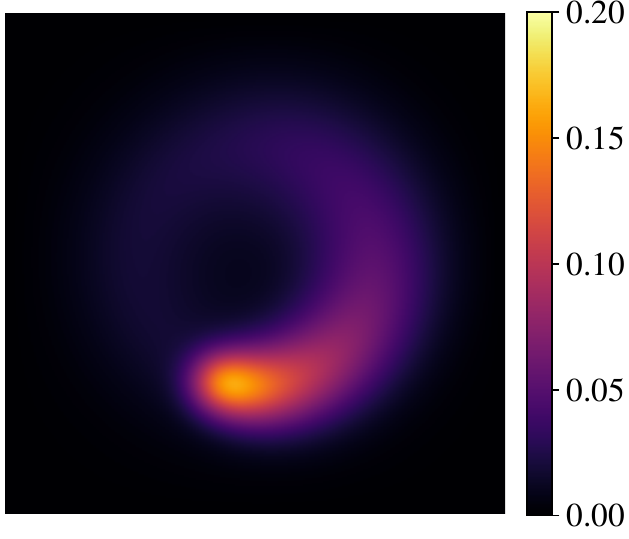}
        \scriptsize
        
        (a) $U(x,y)$.
    \end{minipage}%
    \hfill
    \begin{minipage}{0.228\linewidth}
        \centering
        \includegraphics[height=\figH, clip = true, trim = 16cm 1cm 13.35cm 1.3cm]{images/raddiff8000.pdf}
        \scriptsize
        
        (b) $V(x,y)$.
    \end{minipage}
    \hfill
    \begin{minipage}{0.228\linewidth}
        \centering
        \includegraphics[height=\figH, clip = true, trim = 27.4cm 1cm 2cm 1.3cm]{images/raddiff8000.pdf}
        \scriptsize
        
        (c) $U(x,y) + V(x,y)$.
    \end{minipage}
    \hspace{6mm}
    \caption{Su-Olson radiation diffusion simulation of a orbiting source in circular trajectory with Gaussian amplitude, characterized by unit amplitude and standard deviation $0.4$. Shown are the scalar fields $U(x,y)$, $V(x,y)$, and their combined intensity $U+V$. The computational domain is $x,y \in [-5,5]$. The parameters are $Q_0 = 1$ and $\xi = 0.1$. The simulation is performed on a $512 \times 512$ grid with $\mathrm{C} = 1$ after 8000 time steps.}
    \label{fig:raddiff}


\end{figure}

\begin{figure}[t!]
    \centering

    \def\figH{2.95cm} 
    
    \hspace{6mm}
    \begin{minipage}{0.228\linewidth}
        \centering
        \includegraphics[height=\figH, clip = true, trim = 1.5cm 0.5cm 0.5cm 0.5cm]{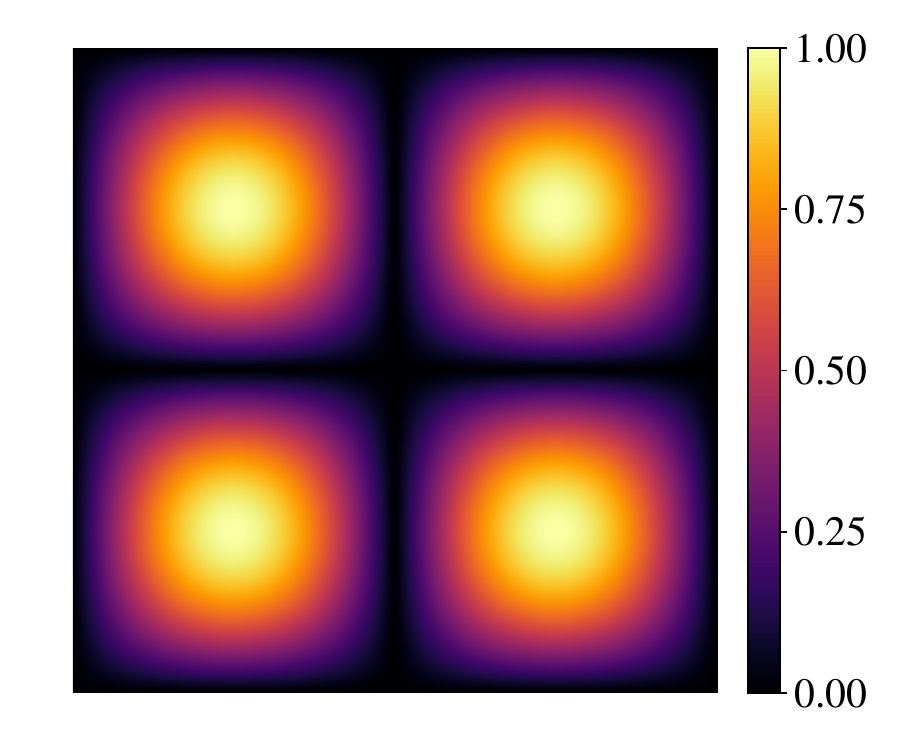}
        \scriptsize
        
        (a) $u(x,y, t_0)$.
    \end{minipage}
    \hfill
    \begin{minipage}{0.228\linewidth}
        \centering
        \includegraphics[height=\figH, clip = true, trim = 1.5cm 0.5cm 0.5cm 0.5cm]{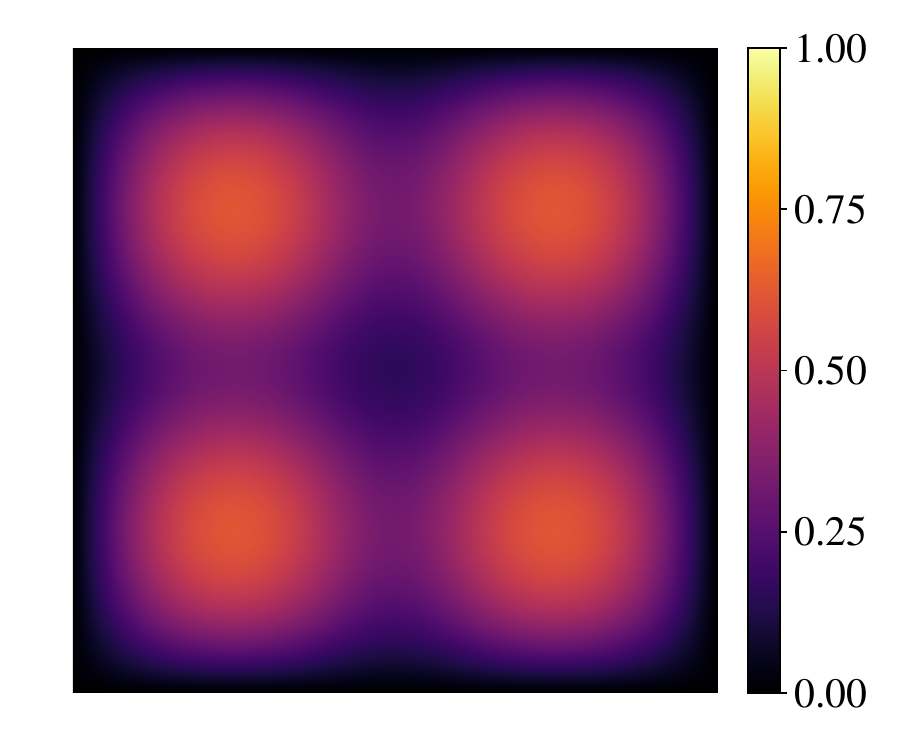}
        \scriptsize
        
        (a) $u(x,y,t_1)$.
    \end{minipage}
    \hfill
    \begin{minipage}{0.228\linewidth}
        \centering
        \includegraphics[height=\figH, clip = true, trim = 1.5cm 0.5cm 0.5cm 0.5cm]{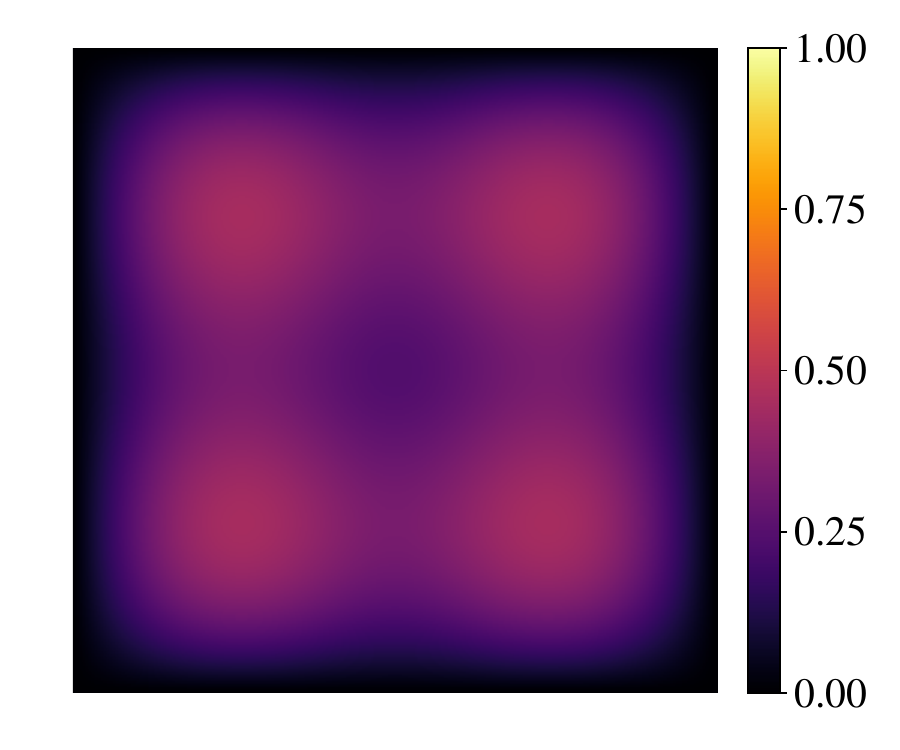}
        \scriptsize
        
        (a) $u(x,y,t_2)$.
    \end{minipage}
    \hspace{6mm}
    \caption{Reaction–diffusion solution on a sinusoidally perturbed grid, illustrating the formation of spatial patterns in $u(x,y)$. The computational domain is $x,y \in [-0.5,0.5]$. The diffusion coefficient is set to $D = 0.01$. The simulation is performed on a $512 \times 512$ grid with $\mathrm{C} = 1$. Results are shown at the initial condition ($t_0$), after 1000 time steps ($t_1$), and after 2000 time steps ($t_2$).}
    \label{fig:reactdiff}

    \vspace{4.5mm}
    
    \hspace{6mm}
    \begin{minipage}{0.228\linewidth}
        \centering
        \includegraphics[height=\figH, clip = true, trim = 0.3cm 0 25cm 18cm]{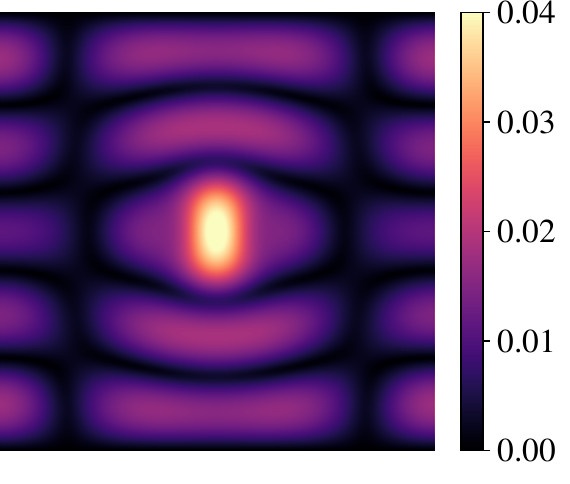}
        \scriptsize
        (a) $|E_x(x,y)|$, $E_x \in \mathbb{C}$.
    \end{minipage}%
    \hfill
    \begin{minipage}{0.228\linewidth}
        \centering
        \includegraphics[height=\figH, clip = true, trim = 12cm 0 12.5cm 18cm]{images/w1.pdf}
        \scriptsize
        (b) $|E_y(x,y)|$, $E_y \in \mathbb{C}$.
    \end{minipage}
    \hfill
    \begin{minipage}{0.228\linewidth}
        \centering
        \includegraphics[height=\figH, clip = true, trim = 24cm 0 0cm 18cm]{images/w1.pdf}
        \scriptsize
        (c) $|\mathbf{E}(x,y)|$, $\mathbf{E} \in \mathbb{C}^2$.
    \end{minipage}
    \hspace{6mm}
    \caption{Symmetric Gaussian source simulation for the time-harmonic Maxwell equations on a $512 \times 512$ grid at normalized frequency $\omega = 18$, with $\chi = 0.05$. The computational domain is $x,y \in [0,1]$ with Perfect Electric Conductor (PEC) boundary conditions, representing a fully enclosed resonant cavity.}
    \label{fig:gauss_sim}
\end{figure}

Before comparing linearization strategies, we verify that the benchmark problems produce the expected physical behavior and that the discretization error is not dominated by the choice of JVP evaluation. Figures~\ref{fig:tgv_sim}--\ref{fig:gauss_sim} show representative solutions from the four benchmark classes: Burgers dynamics, Su--Olson radiation diffusion, reaction--diffusion, and the nonlinear time-harmonic Maxwell problem. These cases illustrate the range of regimes used in the performance study, including advective steepening, coupled radiation--material transport, nonlinear dissipation, and resonant electromagnetic fields.

\begin{figure}[t]
    \centering

    \begin{minipage}{0.3\linewidth}
        \centering
        \includegraphics[width=1\linewidth, clip=true, trim=0cm 0 25.5cm 0]{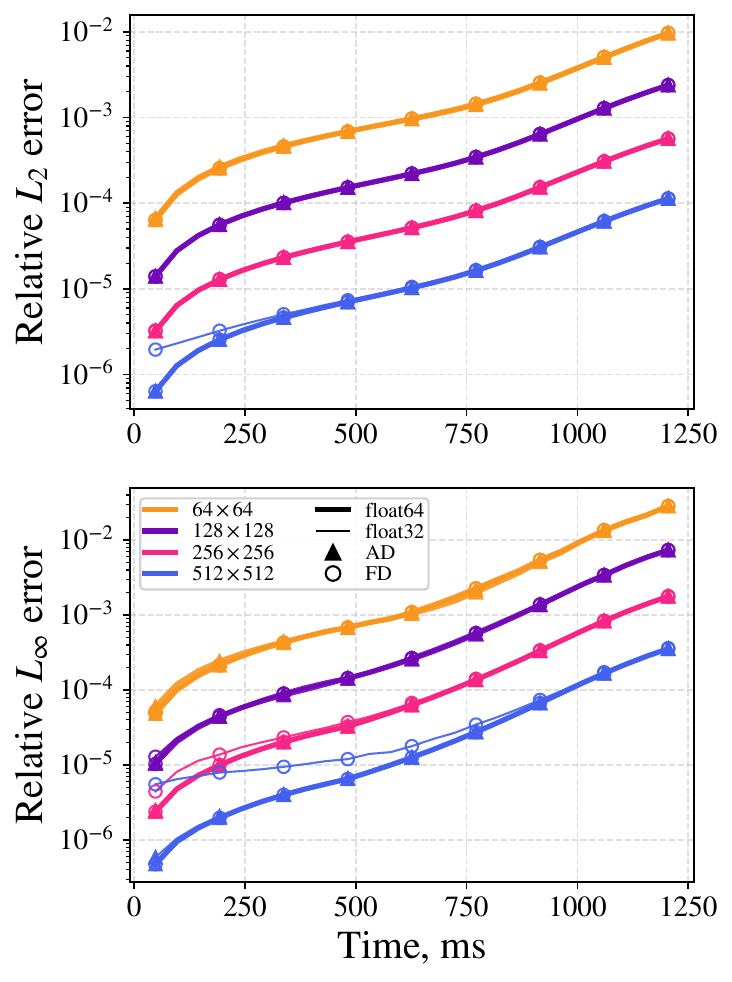}

        \footnotesize
        (a) Error in time
    \end{minipage}
    \hspace{0mm}
    \begin{minipage}{0.292\linewidth}
        \centering
        \includegraphics[width=0.96\linewidth, clip=true, trim=12.3cm 0 14cm 0]{images/combined_plots_new.pdf}

        \footnotesize
        (b) Spatial convergence
    \end{minipage}
    \hfill
    \begin{minipage}{0.34\linewidth}
        \centering
        \includegraphics[width=1\linewidth, clip=true, trim=24cm 0 0cm 0]{images/combined_plots_new.pdf}

        \footnotesize
        (c) Newton convergence
    \end{minipage}

    \caption{Verification and convergence behavior of the JFNK solver. The reference solution is computed on a \(1024\times1024\) grid with AD in FP64. (a) Temporal error evolution for the Burgers Taylor--Green vortex with \(\mathrm{C}=1\). (b) Spatial convergence at final time, showing the expected second-order behavior from the central-difference discretization. AD and FD produce comparable discretization errors, although a precision floor appears for FD in FP32 at the finest grid. (c) Newton convergence for the nonlinear Maxwell--Kerr problem at different frequencies, showing slower convergence near resonant regimes.}
    \label{fig:convergenceOfSolver}
\end{figure}

Figure~\ref{fig:convergenceOfSolver} summarizes the verification results. The Burgers Taylor--Green vortex shows comparable temporal error growth in FP32 and FP64 across grid resolutions. The spatial convergence at final time follows the expected second-order behavior of the central-difference discretization, using a \(1024\times1024\) AD-FP64 solution as reference. A precision floor is visible for FD in FP32 at the finest grid, but the dominant AD/FD differences observed later arise from Krylov convergence rather than from the baseline discretization error.

The Newton convergence curves for the Maxwell--Kerr problem show the increasing difficulty of the nonlinear solve near resonant regimes. Frequencies closer to resonance require more Newton iterations, while non-resonant cases converge more rapidly. This confirms that the benchmark suite contains both relatively benign nonlinear time steps and stiff steady nonlinear solves.

\subsection{JVP cost and Krylov stagnation}
\label{subsec:jvp_krylov_stagnation}
The end-to-end performance gains of AD are not derived from cheaper derivative evaluations. As shown in Table~\ref{tab:jvp_overhead_merged}, AD and FD exhibit comparable per-call costs on both CPU and GPU backends, with AD being slightly more expensive in some cases and slightly faster in others. This indicates that the performance bottleneck is not the cost of evaluating $\mathcal{J}(\mathbf{x}^k)\mathbf{v}_m$ in isolation, but rather lies within the Krylov solver itself. Hence, the dominant factor governing runtime is not the per-iteration Jacobian-vector product, but the number and effectiveness of Krylov iterations required to complete the nonlinear solve.

\begin{table}[b!]
\centering
\caption{Average Jacobian--vector product (JVP) time for a \(256\times256\) Burgers TGV case over 100 time steps. AD and FD have comparable per-call costs, so the dominant runtime differences observed later are mainly due to changes in Krylov and Newton convergence rather than to the cost of an individual JVP.}
\scriptsize
\begin{tabular}{llcccc}
\hline
\textbf{JVP} & \textbf{Precision} & \multicolumn{2}{c}{\textbf{Apple M4}} & \multicolumn{2}{c}{\textbf{NVIDIA A100}} \\
\cline{3-6}
& & \textbf{Avg time (\(\mu\)s)} & \textbf{\% of solver} & \textbf{Avg time (\(\mu\)s)} & \textbf{\% of solver} \\
\hline
FD & FP32 & \(211 \pm 12.0\%\) & \(33.8\%\) & \(256 \pm 8.1\%\) & \(30.7\%\) \\
FD & FP64 & \(396 \pm 25.4\%\) & \(14.6\%\) & \(282 \pm 10.1\%\) & \(14.9\%\) \\
AD & FP32 & \(286 \pm 42.2\%\) & \(14.3\%\) & \(214 \pm 6.4\%\) & \(12.5\%\) \\
AD & FP64 & \(426 \pm 20.1\%\) & \(18.0\%\) & \(228 \pm 9.8\%\) & \(12.8\%\) \\
\hline
\end{tabular}
\label{tab:jvp_overhead_merged}
\end{table}

The main bottleneck appears inside the Krylov solver. Table~\ref{tab:fd_vs_ad_fp32_cpu} shows a representative single-precision Burgers 4-vortex collision (4VC) case in which FD and AD solve the same nonlinear problem with identical Newton and Krylov settings. With FD-JVPs, GMRES repeatedly reaches iteration counts close to the iteration cap, accumulating \(196{,}300\) Krylov iterations. With AD-JVPs, the same solve converges with only 400 Krylov iterations, corresponding to an average of 2 Krylov iterations per Newton step. As a result, the wall-clock solver time decreases from \(99.72\) to \(0.59\) s.

\begin{table}[tb]
\centering
\caption{Representative FP32 Burgers 4VC case comparing FD and AD JVPs with GMRES. Same residual discretization, Newton iteration, line search, Krylov tolerance, and iteration limits are used. Both produce a converged solution with the same number of Newton iterations, but FD repeatedly reaches iteration counts close to the GMRES iteration cap, while AD only requires a few Krylov iterations per Newton loop.}
\scriptsize
\begin{tabular}{lrrrc}
\toprule
\textbf{JVP Method} & \textbf{Solver Time (s)} & \textbf{Newton It.} & \textbf{Total GMRES It.} & \textbf{Converged} \\
\midrule
Finite Difference (FD) & 99.72 & 200 & 196,300 & \checkmark \\
Automatic Diff. (AD)   & 0.59  & 200 & 400     & \checkmark \\
\midrule
& \textbf{169$\times$} & -- & \textbf{491$\times$} & -- \\
\bottomrule
\end{tabular}
\label{tab:fd_vs_ad_fp32_cpu}
\end{table}

This behavior explains why AD can produce large end-to-end speedups even when the cost of an individual AD-JVP is comparable to, or slightly higher than, an FD-JVP. In FP32, the FD perturbation is more sensitive to cancellation and round-off error. The resulting JVPs define a noisy or biased effective operator for GMRES. Since GMRES constructs its Krylov basis from repeated applications of this operator, errors in the JVP can slow residual minimization, trigger restarts, and eventually force the linear solve to stagnate at the iteration cap.

AD avoids this failure mode by differentiating the implemented discrete residual directly. The Krylov solver therefore receives a more consistent matrix-free linearization of the same residual function, without the need to tune a FD perturbation size. The observed speedups are therefore a solver-level effect: AD reduces the number of Krylov iterations and improves the Newton correction, rather than merely accelerating the evaluation of a single derivative.

\subsection{Robustness and performance profiles}
\label{subsec:robustness_profiles}

We next compare solver robustness and efficiency over the full benchmark suite using Dolan--Moré performance profiles. The profiles are built from wall-clock time, with failed runs assigned an infinite performance ratio as described in Section~\ref{subsec:metrics_failure}. Thus, the height of each curve at small \(\tau\) measures efficiency, while the plateau at large \(\tau\) measures the fraction of benchmark configurations successfully solved.

\begin{figure}[t!]
    \centering

    \begin{minipage}{0.99\linewidth}
        \centering
        \includegraphics[width=0.86\linewidth, clip=true, trim=0 0 0cm 0cm]{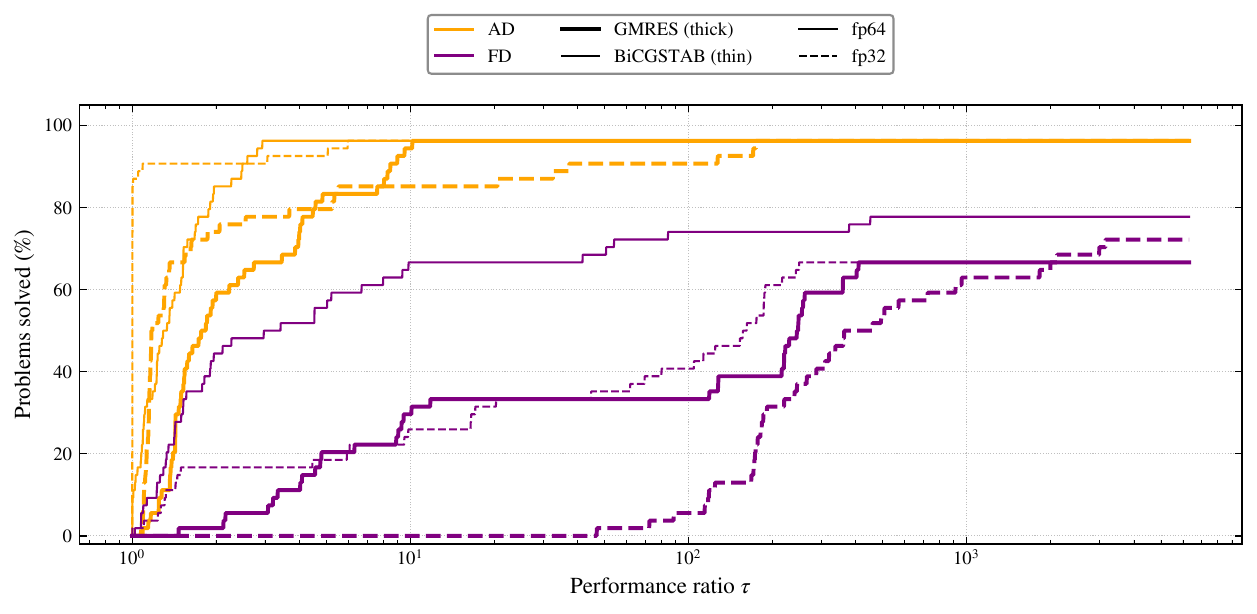}

        \footnotesize
        (a) Dolan--Moré performance profile (CPU)
    \end{minipage}

    \vspace{4mm}

    \begin{minipage}{0.99\linewidth}
        \centering
        \includegraphics[width=0.93\linewidth, clip=true, trim=0 0 0 0.38cm]{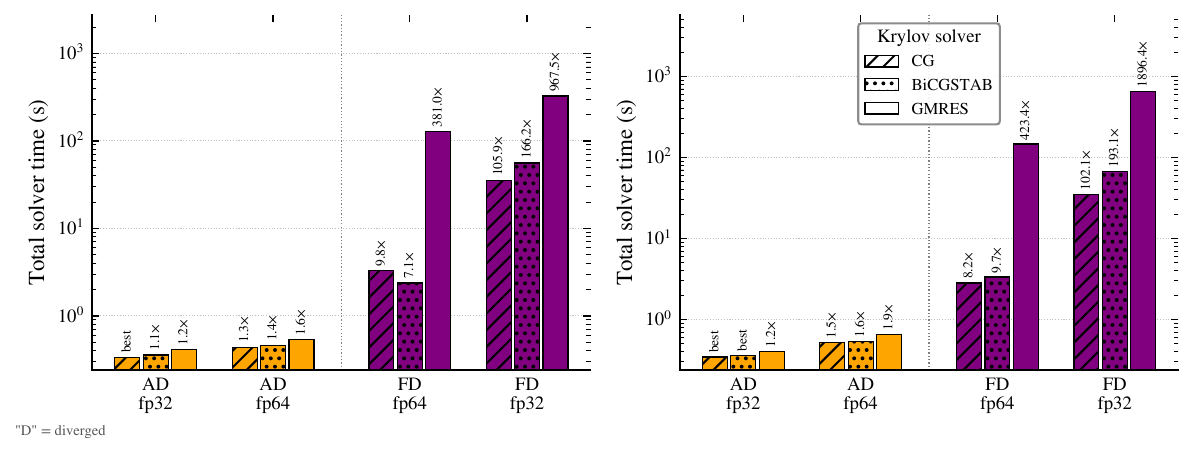}

        \footnotesize
        (b) Reaction--diffusion solver comparison (CPU)
    \end{minipage}

    \caption{
    CPU performance results on the Apple M4 backend. 
    (a) Dolan--Moré performance profiles over the benchmark suite. AD-based JVPs provide higher success rates and lower performance ratios than FD-based JVPs. 
    (b) Reaction--diffusion case on a \(128\times128\) grid with \(D=0.01\), illustrating the benefit of CG when the linearized system is symmetric positive definite.
    }
    \label{fig:m4_performance}
\end{figure}

Figure~\ref{fig:m4_performance}(a) shows the CPU performance profiles. AD-based solvers dominate the upper-left region of the plot, indicating that they are both faster on a larger fraction of problems and more likely to complete the benchmark suite. Among the AD variants, BiCGSTAB is often the fastest option, especially in FP32, because of its low memory footprint and short recurrences. GMRES with AD is slightly less efficient in some cases, but it remains more reliable for difficult nonsymmetric problems. In contrast, FD-based solvers show lower plateaus and much larger performance ratios, particularly in FP32, where FD JVPs can lead to Krylov stagnation.

\begin{figure}[t!]
    \centering

    \begin{minipage}{0.99\linewidth}
        \centering
        \includegraphics[width=0.86\linewidth, clip=true, trim=0cm 0 0cm 0cm]{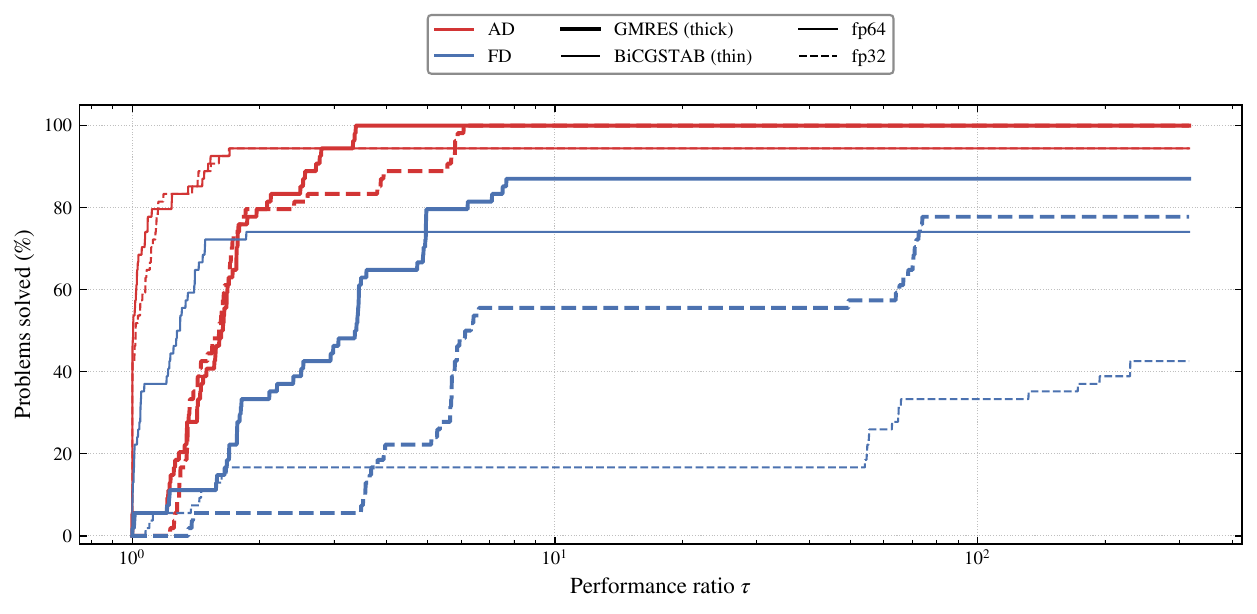}

        \footnotesize
        (a) Dolan--Moré performance profile (GPU)
    \end{minipage}

    \vspace{4mm}

    \begin{minipage}{0.99\linewidth}
        \centering
        \includegraphics[width=0.93\linewidth, clip=true, trim=0 0 0 0.38cm]{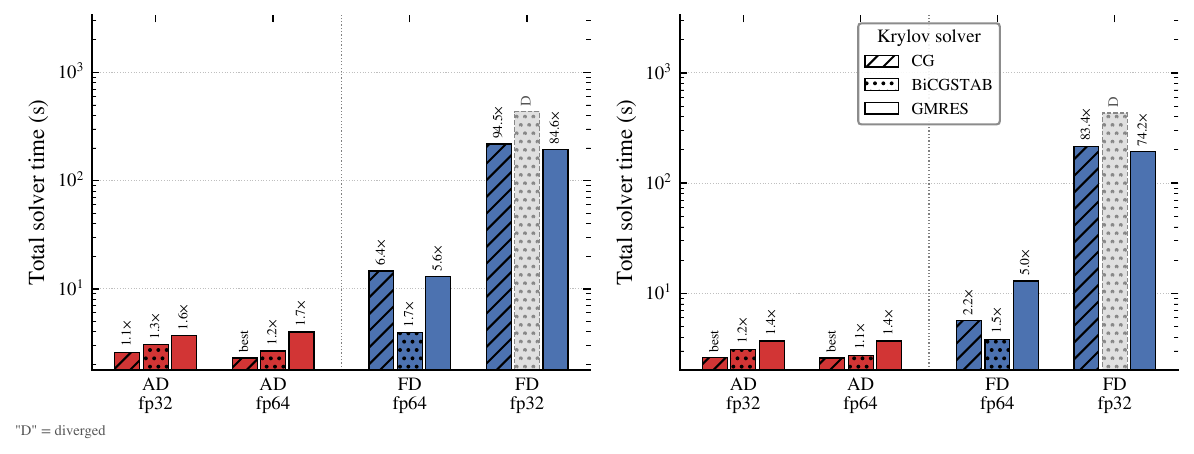}

        \footnotesize
        (b) Reaction--diffusion solver comparison (GPU)
    \end{minipage}

    \caption{
    GPU performance results on the NVIDIA A100 backend. 
    (a) Dolan--Moré performance profiles over the benchmark suite. AD-based solvers achieve higher robustness and better time-to-solution than FD-based solvers. 
    (b) Reaction--diffusion case on a \(512\times512\) grid with \(D=0.01\). For this SPD problem, CG with AD is the most efficient Krylov choice.
    }
    \label{fig:a100_performance}
\end{figure}

The GPU profiles in Figure~\ref{fig:a100_performance}(a) show the same qualitative behavior. AD-based solvers again occupy the most favorable region of the plot. BiCGSTAB with AD gives the best raw performance for many configurations, while GMRES with AD provides the most reliable behavior across the full set of nonlinear problems. FD-based solvers remain substantially less robust, especially in FP32. Although the A100 backend can partially mask large iteration counts through higher throughput, the profile still shows that poor FD linearizations lead to slower and less reliable solves.

The reaction-diffusion panels in Figures~\ref{fig:m4_performance}(b) and~\ref{fig:a100_performance}(b) isolate a case where the linearized system is symmetric positive definite. In this setting, CG is the appropriate Krylov method and gives the lowest runtimes when paired with AD. This confirms that the best Krylov method depends on the algebraic structure of the linearized problem: CG is preferable for SPD systems, BiCGSTAB is often fastest for moderately difficult nonsymmetric systems, and GMRES is more reliable for stiff or ill-conditioned nonsymmetric regimes.

Overall, the performance profiles show that the advantage of AD is not limited to a single problem or backend. Across CPU and GPU runs, AD improves the success rate and reduces the time-to-solution by providing a more reliable matrix-free linearization to the Krylov solver. The largest separation from FD occurs in single precision, where the FD perturbation is most sensitive to round-off and cancellation.

\subsection{Krylov solver selection and failure analysis}
\label{subsec:krylov_failure}
The performance profiles indicate that JVP construction and Krylov solver choice must be considered together. Table~\ref{tab:diverged_configs} summarizes the failed configurations for the CPU and GPU solvers across all 480 benchmark simulations for each backend. Most failures are associated with FD-based JVPs: \(63\) out of \(71\) failures on the CPU backend and \(64\) out of \(70\) failures on the GPU backend. AD-based failures are comparatively rare and are concentrated in configurations where BiCGSTAB is applied to the stiff Maxwell--Kerr dipole problem.

\begin{table}[tb!]
    \centering
    \caption{
    Failed solver configurations. The Su-Olson test cases achieved no failures. Divergence within AD configurations was rare: on the CPU, it was limited to 8 total failures in the Burgers TGV problem, while on the GPU, AD failures were confined to the ill-conditioned Maxwell problem with BiCGSTAB (6 failures). Majority of recorded failures (88.7\% on CPU, 91.4\% on GPU) were driven by FD JVPs. CPU and GPU implementations demonstrated nearly identical global robustness (with 71 and 70 failures over 480 simulations), though direct comparison is limited given different grid resolutions.
    }
    \label{tab:diverged_configs}
    \scriptsize
    \begin{tabular}{llcccc}
        \toprule
        \multirow{2}{*}{\textbf{Problem}} & 
        \multirow{2}{*}{\textbf{Precision / Solver}} & 
        \multicolumn{2}{c}{\textbf{AD failures}} & 
        \multicolumn{2}{c}{\textbf{FD failures}} \\
        \cmidrule(lr){3-4} \cmidrule(lr){5-6}
        & & \textbf{CPU} & \textbf{GPU} & \textbf{CPU} & \textbf{GPU} \\
        \midrule
        \multirow{4}{*}{Burgers TGV} 
        & FP32 BiCGSTAB & 2 & -- & 6 & 6 \\
        & FP32 GMRES    & 2 & -- & 6 & 6 \\
        & FP64 BiCGSTAB & 2 & -- & 4 & 2 \\
        & FP64 GMRES    & 2 & -- & 6 & 2 \\
        \midrule
        \multirow{4}{*}{Burgers 4VC} 
        & FP32 BiCGSTAB & -- & -- & 6 & 4 \\
        & FP32 GMRES    & -- & -- & 3 & -- \\
        & FP64 BiCGSTAB & -- & -- & 3 & 1 \\
        & FP64 GMRES    & -- & -- & 6 & 1 \\
        \midrule
        \multirow{4}{*}{Burgers DSL} 
        & FP32 BiCGSTAB & -- & -- & 6 & 6 \\
        & FP32 GMRES    & -- & -- & 6 & 6 \\
        & FP64 BiCGSTAB & -- & -- & 5 & 4 \\
        & FP64 GMRES    & -- & -- & 6 & 4 \\
        \midrule
        \multirow{2}{*}{Maxwell--Kerr dipole} 
        & FP32 BiCGSTAB & -- & 3 & -- & 3 \\
        & FP64 BiCGSTAB & -- & 3 & -- & 3 \\
        \midrule
        \multirow{2}{*}{Reaction--diffusion Gaussian}
        & FP32 BiCGSTAB & -- & -- & -- & 6 \\
        & FP64 BiCGSTAB & -- & -- & -- & 2 \\
        \midrule
        \multirow{2}{*}{Reaction--diffusion sinusoidal}
        & FP32 BiCGSTAB & -- & -- & -- & 6 \\
        & FP64 BiCGSTAB & -- & -- & -- & 2 \\
        \midrule
        \textbf{Total} & & \textbf{8} & \textbf{6} & \textbf{63} & \textbf{64} \\
        \bottomrule
    \end{tabular}
\end{table}

The table shows that the main robustness difference is not caused by the hardware backend, but by the combination of linearization strategy, precision, and Krylov solver. FD failures are most common in Burgers configurations, especially in FP32, where the FD perturbation is most sensitive to cancellation and round-off. These failures are consistent with the stagnation behavior discussed in Section~\ref{subsec:jvp_krylov_stagnation}: inaccurate JVPs degrade the effective Krylov operator and can prevent the linear residual from decreasing within the iteration limit.

AD substantially reduces this failure mode because the Krylov solver receives the directional derivative of the implemented residual rather than a FD approximation. However, AD does not remove all solver sensitivity. The remaining AD failures occur with BiCGSTAB for the Maxwell dipole case on the GPU. This is consistent with the non-monotone convergence behavior of BiCGSTAB: its short recurrences are efficient, but the method is more vulnerable to breakdown in stiff and ill-conditioned nonsymmetric systems. GMRES with AD is therefore the more reliable choice for the Maxwell--Kerr problem, even when BiCGSTAB is faster on easier nonsymmetric cases.

For the reaction--diffusion problem, the linearized system admits a symmetric positive-definite structure, and CG is the preferred Krylov method. The reaction--diffusion timing panels in Figures~\ref{fig:m4_performance}(b) and~\ref{fig:a100_performance}(b) show that CG with AD gives the lowest runtimes for this class of problems. This confirms that AD improves the quality of the matrix-free linearization, but the Krylov method should still be selected according to the algebraic structure of the linearized system.

Overall, the results suggest the following practical solver choices. For stiff nonsymmetric problems, AD with GMRES provides the most robust configuration. For moderately difficult nonsymmetric problems, AD with BiCGSTAB often gives the best time-to-solution. For symmetric positive-definite problems, AD with CG is preferable. FD-based JVPs are most fragile in FP32 and should be used with caution when the Krylov solve is sensitive to perturbation error.

\section{Conclusions}
\label{sec:conclusion}
This work evaluates FD and AD JVPs in matrix-free JFNK solvers for nonlinear PDEs, covering both time-marching IBVPs and frequency-domain BVPs. The main conclusion is that Jacobian-vector product construction is a critical numerical design choice in JFNK solvers. Although JFNK methods avoid forming the Jacobian, the Krylov solver still acts on a matrix-free representation of the local linearized residual. When this representation is obtained by finite differences, the perturbation parameter introduces sensitivity to residual scaling, cancellation, and round-off error. This sensitivity is particularly important in FP32, where the useful range of perturbation sizes is narrower. In the tested problems, inaccurate FD JVPs degraded the Krylov operator, increased iteration counts, and in some cases prevented the nonlinear solve from converging.

Forward-mode AD mitigates this failure mode by differentiating the implemented discrete residual directly. The resulting JVP improve the quality of the Krylov subproblem, leading to fewer Krylov iterations, more reliable Newton convergence, and much lower failure rates. The most significant differences are observed in single precision, with a clear separation appearing in the Dolan--Mor\'{e} performance profiles. All AD configurations surpassed a 95\% completion rate, whereas FD performance dropped as low as 64\% on CPU and 42\% on GPU. Speedups exceed three orders of magnitude for the CPU solver and two orders of magnitude for the GPU solver in completed simulations. For a representative converged FP32 Burgers case, AD achieved a 169$\times$ speedup over FD by reducing GMRES iterations by 491$\times$ for the same number of Newton iterations.

These gains were not caused by a lower cost per Jacobian-vector product. In the present implementation, AD and FD had comparable per-call costs. Instead, the improvement came from reducing the number of Krylov and Newton iterations required by the full nonlinear solve.

The performance profiles further demonstrate that the benefits of AD are not limited to a single backend. On both CPU and GPU, AD-based solvers achieved higher success rates and better time-to-solution than FD-based solvers. The GPU backend can partially hide inflated iteration counts through higher throughput, but it does not remove the underlying algorithmic inefficiency caused by poor FD linearizations. Thus, the advantage of AD should be interpreted primarily as an improvement in solver robustness, with performance gains following from more reliable Krylov convergence.

The choice of Krylov method remains problem dependent. For nonsymmetric problems, BiCGSTAB with AD was often the fastest option when the linearized systems were moderately stiff. However, its non-monotone convergence behavior made it less reliable for the stiff Maxwell--Kerr dipole cases. GMRES with AD was more robust in these regimes, although sometimes at higher cost, as expected. For reaction-diffusion problems with symmetric positive-definite linearized systems, CG with AD was the most efficient choice. These results indicate that AD improves the matrix-free linearization, but the Krylov solver should still be selected according to the algebraic structure and stiffness of the problem.

The conclusions should also be interpreted within the scope of the residuals considered here. AD differentiates the implemented discrete residual, not the continuous PDE. For smooth and deterministic residual evaluations, this provides a consistent matrix-free linearization and avoids the perturbation-size sensitivity of finite differences. If the residual contains nonsmooth limiters, adaptive discontinuities, stochastic components, or noisy embedded solvers, AD may differentiate algorithmic artifacts. In such cases, carefully chosen finite differences or modified linearization strategies may still be useful. The present results identify the regime where AD is most beneficial: smooth finite-precision JFNK residuals for which FD perturbation error is the dominant source of Krylov degradation.

Overall, this study demonstrates that robust matrix-free JFNK performance depends not only on the choice of Krylov solver or hardware backend, but fundamentally on the construction of the Jacobian-vector products. For the nonlinear PDE benchmarks examined, forward-mode AD provides a more reliable finite-precision linearization than finite differences, particularly within FP32 and stiff nonlinear regimes.

\section*{Acknowledgments}

\noindent This work has received funding from the Swedish Research Council’s Research Environment grant (SEE-6GIA 2024-06482). We thank the CuPy developers for their support during our implementation of the CuPy BiCGSTAB Krylov solver.

\appendix

\section{Numerical Treatment of Test Cases} \label{appdx:testcases}

\setcounter{figure}{0}

\noindent We evaluate the JFNK solver across two distinct families of PDEs: the time-marching initial boundary value problems (IBVPs) and boundary value problems (BVPs). 

Time-marching systems are generally well-conditioned for implicit solvers. The discretization of the time derivative introduces a diagonally dominant mass matrix (scaled by $1/\Delta t$) into the Jacobian, and the converged solution from the previous time step provides a highly accurate initial guess within the Newton quadratic convergence radius. Conversely, BVPs seek a global field distribution coming from a steady-state or frequency domain problem. Lacking a time derivative, the Jacobian is dictated entirely by high-order spatial operators and highly localized nonlinearities, causing severe stiffness, ill-conditioned linear systems. Furthermore, boundary value problems (BVPs) require initialization from an approximate initial guess, placing significant strain on the global convergence properties of the Newton method and its associated line search. In addition, their solutions are complex-valued fields, which effectively doubles the memory requirements during the computation.

\subsection{Burgers' Equation}

\noindent The 2D transient Burgers' equation, driven by convective nonlinearity, is given by:
\begin{equation}
    \frac{\partial \mathbf{U}}{\partial t} + (\mathbf{U} \cdot \nabla)\mathbf{U} = \nu \nabla^2 \mathbf{U}
\end{equation}
where $\mathbf{U} = (u, v)^T$ represents the velocity field and $\nu$ denotes the kinematic viscosity. 

\paragraph{Discretization}
Defining discrete second-order central difference operators for advection, $\mathrm{Adv}(\mathbf{U}) = u D_x \mathbf{U} + v D_y \mathbf{U}$, and diffusion, $\mathrm{Lap}(\mathbf{U}) = D_{xx} \mathbf{U} + D_{yy} \mathbf{U}$, the nonlinear residual at Newton iteration $k$ (targeting time level $n+1$) is expressed in vector form as:
\begin{equation}
    \mathbf{F}(\mathbf{U}^k, \mathbf{U}^{n}) = \mathbf{U}^k - \mathbf{U}^{n} + \frac{\Delta t}{2} \left[ \mathrm{Adv}(\mathbf{U}^k) - \nu \, \mathrm{Lap}(\mathbf{U}^k) + \mathrm{Adv}(\mathbf{U}^{n}) - \nu \, \mathrm{Lap}(\mathbf{U}^{n}) \right]
\end{equation}
The time step is constrained adaptively via Courant–Friedrichs–Lewy (CFL) condition:
\begin{equation}
    \Delta t = C \cdot \min\!\left( \frac{\Delta x}{\max|u|},\; \frac{\Delta y}{\max|v|},\; \left(2\nu\!\left(\frac{1}{\Delta x^2} + \frac{1}{\Delta y^2}\right)\right)^{-1} \right)
\end{equation}

\paragraph{Initial and Boundary Conditions}
The solver is evaluated against three fully periodic benchmark configurations:
\begin{itemize}
    \item {Taylor-Green Vortex (TGV):} An analytical baseline tracking exponential kinetic energy decay, initialized as $u = \sin(x)\cos(y)$ and $v = -\cos(x)\sin(y)$.
    \item {Double Shear Layer (DSL):} A rigorous nonlinear stress test driven by Kelvin-Helmholtz instabilities ($\rho = 30$, $\delta = 0.05$):
    \begin{equation}
        u(x, y) = \begin{cases} \tanh\!\left(\rho \left(y - \frac{\pi}{2}\right)\right) & y \leq \pi \\ \tanh\!\left(\rho \left(\frac{3\pi}{2} - y\right)\right) & y > \pi \end{cases}, \quad v(x, y) = \delta \sin(x)
    \end{equation}
    \item {4-Vortex Collision (4VC):} Evaluates line search robustness during violent flow reorganization via four alternating-sign Gaussian vortices ($R=0.5$):
    \begin{equation}
        \mathbf{U}(x,y) = \sum_{i=1}^{4} \Gamma_i \begin{pmatrix} -(y - c_{y,i}) \\ (x - c_{x,i}) \end{pmatrix} e^{-r_i^2 / R^2}
    \end{equation}
    where $r_i^2 = (x - c_{x,i})^2 + (y - c_{y,i})^2$.
\end{itemize}

\subsection{Non-Equilibrium Radiative Diffusion (Su-Olson)}

\noindent The Su-Olson problem models dimensionless non-equilibrium radiation transport, coupling a diffusing radiation energy field $U$ to a localized material energy field $V$:
\begin{equation}
    \frac{\partial U}{\partial t} = \frac{1}{3} \nabla^2 U - (U - V) + Q(\mathbf{x}, t), \quad \frac{\partial V}{\partial t} = \xi (U - V)
\end{equation}
In the physical regime where $\xi \ll 1$, the disparity in timescales between rapid radiation diffusion and slow localized material absorption renders the system highly stiff.

\paragraph{Discretization}
The coupled discrete residual under Crank--Nicolson integration is formulated as:
\begin{equation}
    \begin{aligned}
        F_U &= U^k - U^n - \frac{\Delta t}{2} \left[ \frac{1}{3}\mathrm{Lap} (U^k + U^n) - (U^k + U^n) + (V^k + V^n) + 2Q \right] \\
        F_V &= V^k - V^n - \frac{\Delta t}{2} \xi \left[ (U^k + U^n) - (V^k + V^n) \right]
    \end{aligned}
\end{equation}
subject to the multi-physics CFL constraint:
\begin{equation}
    \Delta t = \min\left( C \frac{\min(\Delta x, \Delta y)^2}{D},\; \frac{C}{\xi + \delta} \right)
\end{equation}

\paragraph{Configurations}
We explore two spatial configurations. The classic 1D Su-Olson profile maps to a 2D grid with periodic $y$-boundaries, applying a cold start ($U=V=0$) driven by a central, stationary box source $Q_0$. The dynamic 2D configuration stresses fully enclosed spatial coupling within strict Dirichlet boundaries, initializing decaying concentric rings and an orbiting Gaussian source:
\begin{equation}
    Q(x, y, t) = Q_0 \exp\!\left(-\frac{(x - R\cos(\omega t))^2 + (y - R\sin(\omega t))^2}{2\sigma^2}\right)
\end{equation}

\subsection{Nonlinear Reaction-Diffusion}

\noindent This equation models a scalar concentration field $u$ subject to a volumetric nonlinear depletion sink:
\begin{equation}
    \frac{\partial u}{\partial t} = D \nabla^2 u - u^3
\end{equation}
Because the analytical derivative of the sink is strictly non-positive ($f'(u) = -3u^2 \leq 0$) and the discrete Laplacian is negative-definite, the Jacobian matrix, $\mathcal{J} = \mathbf{I} - \frac{\Delta t}{2} D \mathbf{L} + \frac{3\Delta t}{2} (u^k)^2 \mathbf{I}$, is inherently Symmetric Positive Definite (SPD). This property guarantees Conjugate Gradient (CG) convergence, allowing for direct comparisons against GMRES and BiCGSTAB both on CPU and GPUs. 

The domain is entirely bounded by homogeneous Dirichlet conditions. We initialize the system with three concentration profiles: a single Gaussian, a Multi-Gaussian with four interacting peaks, and a high-gradient Sinusoidal field $u(x,y) = |\sin(k_x \pi x) \sin(k_y \pi y)|$ to evaluate rapid gradient decay. The system is discretized analogously to previous cases, with CFL condition:
\begin{equation}
    \Delta t = C  \Delta x^2 / D
\end{equation}

\subsection{Time-Harmonic Maxwell's Equations in Kerr Media}

\noindent The time-harmonic Maxwell's wave equation in a nonlinear medium is defined as:
\begin{equation}
    \nabla \times \nabla \times \mathbf{E} - \omega^2 \mu_0 \, \varepsilon(\mathbf{E}) \, \mathbf{E} = i\omega\mu_0 \mathbf{J}
\end{equation}

\begin{figure}[t!]
    \centering
    \includegraphics[width=0.75\linewidth]{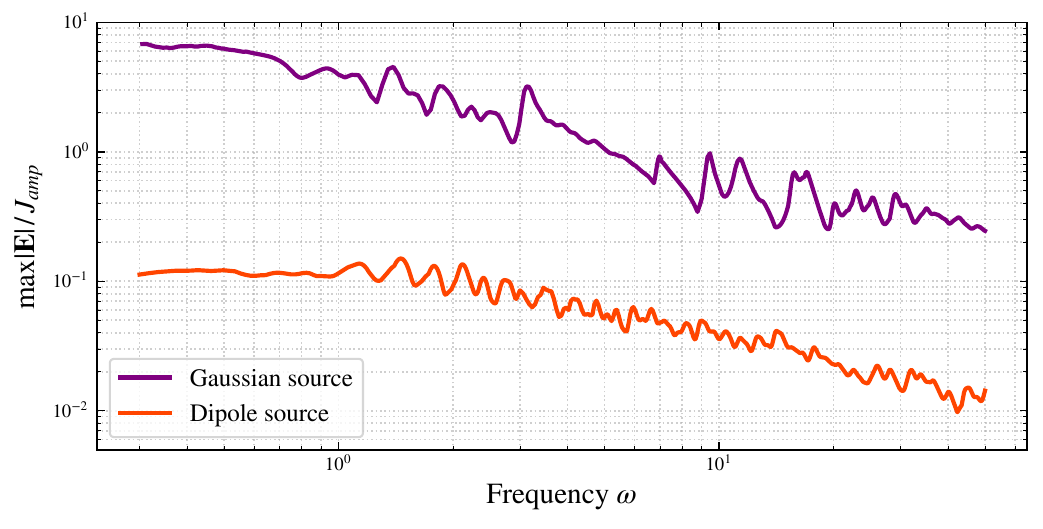}
    \caption{
    Resonance plot for the Maxwell equations in a Kerr medium with $\chi = 0.05$, normalized by the magnitude of the current.
    }
    \label{fig:kerr_resonance}
\end{figure}

For the 2D Transverse Electric (TE) formulation ($\mathbf{E} = (E_x, E_y, 0)^T$), we model a Kerr-type medium with complex dielectric loss:
\begin{equation}
    \varepsilon(\mathbf{E}) = \varepsilon_0 (1 - 0.05i)\!\left(1 + \chi \sqrt{|E_x|^2 + |E_y|^2}\right)
\end{equation}
where $\chi = 0.05$, while $\varepsilon_0 = 1.0$ and $\mu_0 = 1.0$. This localized material dependence fundamentally cross-couples the permittivity to the unknown field solution, rendering the system exceptionally ill-conditioned, especially at resonant frequencies (see Fig.~\ref{fig:kerr_resonance}).

\paragraph{Discretization and Initialization}
Applying second-order central differences to the curl-curl operator yields fully coupled residuals $F_x$ and $F_y$. PEC boundaries are enforced explicitly via $F = E_{\text{tangential}} = 0$. To guarantee Newton-Krylov stability, the initial field guess is computed using the Born approximation, solving the naturally linearized wave equation at $\mathbf{E} = \mathbf{0}$ where $\varepsilon = \varepsilon_0 (1 - 0.05i)$.

\paragraph{Preconditioning}
To accelerate inner Krylov solvers near structural resonances ($\omega$), we deploy a matrix-free, block-diagonal Complex Shifted Laplacian (CSLP) preconditioner. The local permittivity is artificially damped ($\varepsilon_{\text{precond}} = \varepsilon(\mathbf{E}) (1 - 0.5i)$) to shift the operator's eigenvalues into the right half-plane. The decoupled diagonal components are:
\begin{equation}
    M_{xx} = \frac{2}{\Delta y^2} - \omega^2 \mu_0 \, \varepsilon_{\text{precond}}, \quad M_{yy} = \frac{2}{\Delta x^2} - \omega^2 \mu_0 \, \varepsilon_{\text{precond}}
\end{equation}

\paragraph{Source Currents}
The frequency response is mapped via two applied source currents ($\mathbf{J}$):
\begin{itemize}
    \item {Symmetric Gaussian:} $J_{x} = \exp(-((x-x_0)^2+(y-y_i)^2)/(2\sigma^2))$ and $J_y = 0$.
    \item {Antisymmetric Dipole:}
    \begin{equation}
        J_x = \exp\left(-\frac{(x-x_0)^2+(y-y_1)^2}{2\sigma^2}\right) - \exp\left(-\frac{(x-x_0)^2+(y-y_2)^2}{2\sigma^2}\right)
    \end{equation} 
    and $J_y = 0$.
\end{itemize}

\bibliographystyle{elsarticle-num} 
\bibliography{bibliography}







\end{document}